\documentclass[prd,twocolumn, superscriptaddress]{revtex4}

\usepackage{amsmath,amsthm,amssymb}
\usepackage{graphics,graphicx}
\usepackage{epsfig}
\usepackage{color}
\makeatletter
\@addtoreset{equation}{section}
\makeatother

\begin{document}
\title{Bounds on Tensor wave and Twisted Inflation}
\author{Sudhakar Panda} \email{panda@mri.ernet.in}\affiliation{Harish-Chandra Research Institute, Allahabad-211019, India.}
\author{M. Sami} \email{sami@iucaa.ernet.in}\affiliation{Center for Theoretical Physics, Jamia Millia Islamia, Jamia Nagar, New Delhi-110092, India.}
\author{John Ward} \email{jwa@uvic.ca} \affiliation{ Department of Physics and Astronomy, University of Victoria, Victoria, BC, V8P 5C2, Canada.  }

\date{July 2010}
\begin{abstract}
We study the bounds on tensor wave  in a class of twisted inflation models, where $D(4+2k)$-branes are wrapped on cycles in the compact manifold
and wrap the KK-direction in the corresponding effective field theory. While the lower bound is found to be analogous to  that in Type IIB models of brane inflation,  the upper bound turns out to be
significantly different. This is argued for a range of values for the parameter $g_sM$  satisfying the self-consistency relation and the WMAP
data. Further, we observe that the wrapped $D8$-brane appears to be the most attractive from a cosmological perspective.
\end{abstract}
\maketitle
\section{Introduction}
Brane inflation, in a warped compactification , is one of the most appealing open-string inflationary models - bypassing the $\eta$ problem
by considering the relativistic limit of the theory \cite{Silverstein:2003hf}. 
Inflation is realized when the 
scalar potential dominates the kinetic terms even
for relativistic rolling, because the high degree of warping tends to suppress the kinetic contribution. Because of the inherently non-linear
nature of the theory, there is mixing between different modes of the inflaton field. One of the consequences is that there is a non-zero
bispectrum, suggesting that the theory may predict large levels of non-Gaussian perturbations \cite{Lidsey:2006ia}. 
Moreover for large field inflation,
the distance over which the scalar field can roll $\Delta \phi$ must be bounded by the size of the extra dimensions. This in turn can
generate an upper bound on the tensor-scalar ratio using application of the Lyth bound \cite{Lyth:1996im}. In the case of a single $D3$-brane, the two
bounds on $r$ are inconsistent. The maximal predicted value for $r$ is lower than the minimum allowed value \cite{Lidsey:2007gq} and
therefore the validity of the theory is highly questionable.

In the context of type IIB theory, it was noted that $\Delta \phi$ can increase for branes wrapped on cycles within the compact space \cite{Kobayashi:2007hm}. 
As such this helps to relieve the fine-tuning necessary for DBI inflation. Ultimately however, such configurations are subject to important 
back-reaction effects which severely restricts the parameter space of solutions.
The Type IIA theory has not been investigated in as much detail. A recent viable model dubbed Twisted Inflation \cite{Davis:2010it}, 
has been proposed in the non-relativistic limit. One can ask whether the relativistic limit admits similar bounds on the tensor-scalar ratio, and if so; how
are they modified when one considers a simple scenario where higher dimensional branes are wrapped on internal cycles. This is what
we consider in this note. We begin in Section I by briefly discussing the Twisted Inflation scenario , and extending it to a special class of wrapped
brane solutions. Whilst we do not generally know the full scalar potential in such theories \footnote{The wrapped $D8$-brane theory requires massive type IIA
supergravity. One can, in principle, compute the one-loop corrections to the effective potential, however we believe that this correction simply shifts
the value of $V_0$ since the wrapping changes the effective tension of the brane. Work on this direction is in progress.} this is not
necessary to derive bounds on $r$ in the relativistic limit. 
In Section II we then determine the various cosmological constraints on the model and conclude with a brief discussion.
\section{Strong coupling limit}
We consider the strong ('t Hooft) coupling limit of the field theory model, which has a dual gravity description in terms of type IIA supergravity provided
the coupling satisfies $1 << \lambda << M^{2/3}$. If the coupling is larger, then the correct dual description becomes M-theory.
Specifically we consider a background generated by $M$ coincident $D4$-branes, with a compact direction denoted by $x_4$ - which is the KK-direction in the field
theory - and has radius given by $2 \pi R$ \cite{Davis:2010it}. The relevant background parameters are defined below;
\begin{eqnarray}
ds^2 &=& \left(\frac{\rho}{L}\right)^{3/2} \left(\eta_{\mu \nu} dx^{\mu}dx^{\nu} + f(\rho)dx_4^2 \right) \nonumber \\
&+& \left (\frac{L}{\rho}\right)^{3/2} \left(\frac{d\rho^2}{f(\rho)} + \rho^2 d\Omega_4^2 \right) \nonumber \\
f(\rho) &=& 1-\frac{\rho_0^3}{\rho^3} \nonumber \\
e^{\phi} &=& \left( \frac{\rho}{L}\right)^{3/4} \\
C_{(5)} &=& \left( \frac{\rho}{L}\right)^3 \sqrt{f(\rho)} dx_0 \wedge \ldots \wedge dx_4 \nonumber
\end{eqnarray}
where $\rho_0$ is a cut-off which in the dual theory gives rise to confinement, analogous to the Klebanov-Strassler result \cite{Klebanov:2000hb}.
In \cite{Davis:2010it}, a $D4$-brane is wrapped on the compact direction but is dynamical along the throat direction
parameterised by $\rho$. This is essentially a way of realizing DBI inflation \cite{Silverstein:2003hf} in type IIA string theory. 
As in that case, we should consider extensions of the model which generalize the results to wrapped configurations - since
these avoid certain pathologies of the single (flat) brane models \cite{Becker:2007ui, Ward:2007gs, Mukohyama:2007ig, Huston:2008ku, Kobayashi:2007hm}. 
As such we will consider a simple extension of the above model to
$D(4+2k)$-branes (where $k=0, 1, 2$) which are wrapped on a circle ( the KK-direction ), a $\Sigma_2$ or $\Sigma_4$ cycles in the 
compact four-manifold, but allow for dynamical evolution along the $\rho$ direction. We will not turn on gauge fields in this model
to keep things simple. One must note that the dilaton is now non-trivial and therefore the gauge field coupling in Einstein frame will
be different to that in string frame. Upon calculation of the induced metric, we can write the DBI action for the wrapped brane as follows:
\begin{eqnarray}
S &=& -T_{4+2k} \int d^4 \xi 2\pi R \left(\frac{\sqrt{f} \rho^3}{L^3}\right) \sqrt{1+ L^3 \frac{\partial_\mu \rho \partial_\nu \rho \eta^{\mu \nu}}{\rho^3 f(\rho)}} \nonumber \\
&\times& \int d^{2k}\xi \sqrt{\rm{det}G_{kl}}
\end{eqnarray}
where we have the wrapped brane on the $2k$-cycle, and denoted the transverse embedding by 
\begin{equation}
G_{kl} = \left( \frac{L}{\rho}\right)^{3/2} \rho^2 g^{(4)}_{mn}\partial_k y^m \partial_l y^n
\end{equation}
where $g^{(4)}_{mn}$ is the metric on the four-manifold. Note that we have already integrated out the KK-direction in order to
simplify the form of the action. Let us now define the following variables which brings the above theory to the canonical
form of the action for DBI inflation:
\begin{eqnarray}
T(\phi) &=& T_{4+2k}2\pi R\left(\sqrt{f(\rho)}\frac{\rho^3}{L^3} \right)\int d^{2k}\xi\sqrt{\rm{det}G_{kl}} \nonumber \\
d\phi &=& \sqrt{2 \pi R T_{4+2k}}\left(\int d^{2k}\xi \sqrt{\rm{det}G_{kl}} \right)^{1/2} d\rho
\end{eqnarray}
and with the inclusion of the Chern-Simons term we can then write the BPS action in the following (familiar) form
\begin{equation}
S = - \int d^4 \xi T(\phi) \sqrt{1+\partial_\mu \phi \partial_\nu \phi \eta^{\mu \nu} T(\phi)^{-1}}+\int d^4 \xi T(\phi).
\end{equation}
At this stage we now want to couple this to the usual Einstein-Hilbert action. This can be achieved by promoting the
four-dimensional Minkowski metric to FRW form with vanishing spatial curvature. We can also couple a non-trivial scalar potential to this theory (but our analysis below is independent of the form of the scalar potential) and the
resulting action takes the form
\begin{eqnarray}
S = \int d^4 \xi \sqrt{-g}\left(\frac{M_p^2}{2}\mathcal{R} - T(\phi)\left(\gamma^{-1}-1 \right) -V(\phi)\right)
\end{eqnarray}
where we have introduced the relativistic notation $\gamma^{-1} = \sqrt{1 - \dot{\phi}^2 T^{-1}}$
as is standard in such models. Twisted inflation was considered in the non-relativistic limit where $\gamma \sim 1$. We will
consider the full action and make comments about the non-relativistic limit when appropriate - but the primary focus is the 
relativistic limit where we can neglect the coupling to the RR-form field.
\section{Tensor constraints}
The action has now been written in canonical DBI formulation, therefore we can immediately write down the cosmological
parameters that characterize the dynamic evolution in the Hamilton-Jacobi formalism \cite{Kobayashi:2007hm};
\begin{equation}
\epsilon = \frac{2 M_p^2}{\gamma} \left(\frac{H'}{H} \right)^2,\hspace{0.1cm} \eta = \frac{2 M_p^2 H''}{\gamma H}, \hspace{0.1cm} s = \frac{2 M_p^2 \gamma'H'}{\gamma^2 H}
\end{equation}
where $H$ is the Hubble parameter, given in terms of the energy density of the scalar field sector. Primes denote derivatives with respect to the 
inflaton field ($\phi$). The above slow roll parameters must be less than unity for inflation to occur. We can then write the relevant
power spectra and scalar indices at leading order
\begin{eqnarray}
P_s^2 &=& \frac{H^2 \gamma}{8 \pi^2 M_p^2 \epsilon}, \hspace{1cm} P_t^2 = \frac{2 H^2}{\pi^2 M_p^2} \nonumber \\
n_s-1 &=& 2\eta-4\epsilon-2s, \hspace{0.51cm} n_t = -2\epsilon \\
r &=& \frac{16 \epsilon}{\gamma}, \hspace{1.7cm} f_{NL} = \frac{1}{3} \left(\gamma^2-1 \right)
\end{eqnarray}
note that the non-Gaussianity parameter $f_{NL}$ increases with $\gamma$ and that the tensor to scalar ratio $r$ is suppressed at large $\gamma$.
In the models based on $D3$-branes, it was suggested that this could lead to a detectable signal for string theory. However various
inconsistencies are present in the theory which means that the cosmological observables are only valid for small $\gamma$ which leads to
models that are indistinguishable from canonical slow-roll theories \cite{Lidsey:2007gq}. It should also go without saying that the above expressions
should be evaluated at horizon crossing $k = a \gamma H$, although since the tensor mode horizon remains at $k = aH$. This is 
because the scalar potential, which we have inserted by hand, dominates the Hubble factor even for relativistic velocities. This
ensures that the tensor power spectrum is insensitive to $\gamma$, however the observable $r$ does depend on this term. Under
the assumption that $\gamma >> 1$, we see that there is a potential to generate large non-Gaussianity. A useful relation
(using this assumption) is the following \cite{Kobayashi:2007hm}
\begin{equation}\label{eq:identity}
T(\phi) \sim \frac{\pi^2 r^2 M_p^4}{16} P_s^2 \left(1 + \frac{1}{3 f_{NL}} \right)
\end{equation}
which relates the tensor-scalar ratio to the scalar power spectrum and the non-Gaussian parameter.

The four-dimensional action we work with requires that there are six compact dimensions. This is because the four-dimensional (effective) Planck
mass is a function of these dimensions through the usual formula
\begin{equation}\label{eq:Planck_mass}
M_p^2 = \frac{V_6}{\kappa_{10}^2}
\end{equation}
where $V_6$ is the internal volume and $\kappa_{10}^2$ is the ten-dimensional Newton constant. 
Therefore if the six-manifold is non-compact this results in an infinite (effective) Planck scale. We will assume that the supergravity background
is compact, this could be achieved by gluing the UV region to another manifold. The six-dimensional (warped) volume is given by the product
of the four-manifold, the KK-direction and the throat geometry, which can be estimated to be
\begin{equation}
V_6 \sim \int d\rho 2\pi R L^{3/2} \rho^{5/2} \Sigma_4
\end{equation}
where $\Sigma_4$ is the volume of the four-manifold. Given this solution, we can estimate the upper bound on the tensor to scalar ratio
using the analysis in \cite{Lidsey:2007gq, Lidsey:2006ia},\footnote{An alternate analysis was proposed in\cite{Baumann:2006cd}, 
however we are interested in the most conservative of the two bounds}, 
by assuming that the following bounds are satisfied for large field (UV) inflation, whereby the
brane moves towards the tip of the throat;
\begin{equation}
\rho_{*} >> |\Delta \rho_{*}|, \hspace{1cm} V_6 >> |\Delta V_{6,*}|
\end{equation}
we can estimate the change in volume through the relation $\Delta V_6 \sim \Sigma_4 2 \pi R L^{3/2} \Delta \rho_{*}^{7/2}$. 
We also denote the unit volume of the wrapped cycle by
\begin{equation}
v_{2k} = \int d^2 \xi \sqrt{{\rm det} g_{mn} \partial_k y^m \partial_l y^n}
\end{equation}
and note that when $k=0$ we must use $v_{2k} = 1$. The modified canonical transformations become
\begin{eqnarray}
T(\phi) &=& T_{4+2k}2\pi R\left(\sqrt{f(\rho)}\frac{\rho^3}{L^3} \right)\int d^{2k}\xi\sqrt{\rm{det}G_{kl}} \nonumber \\
d\phi &=& \sqrt{2 \pi R T_{4+2k}}\left(\int d^{2k}\xi \sqrt{\rm{det}G_{kl}} \right)^{1/2} d\rho.
\end{eqnarray}
The Lyth bound \cite{Lyth:1996im} in such a model takes the form
\begin{equation}
\left( \frac{\Delta \phi_{*}}{M_p}\right)^2 = \frac{r}{8} (\Delta N_{*})^2
\end{equation}
where $\Delta N_{*}$ corresponds to the amount of \emph{visible} inflation. The Lyth bound yields the following solution
\begin{equation}
\left( \frac{\Delta \phi_{*}}{M_p}\right)^2 < \frac{\kappa_{10}^2 T_{4+2k}v_{2k}}{\Sigma_4 L^{3/2}} \frac{(L^3 \rho_{*})^{k/2}}{(\Delta \rho_{*})^{3/2}}
\end{equation}
where we have used the definition of the compact volume. Note that we can keep terms in the numerator proportional to $\rho_{*}$, since the aim is
to bound $r$ from above. Now using (\ref{eq:identity}) we can solve for the brane tension in the expression above, and we find the solution
\begin{equation}
r < \frac{\kappa_{10}^2 R T_{4+2k}^2 v_{2k}^2 \pi^3 (L^3 \rho_{*})^k L}{64 \Sigma_4 (\Delta N_{*})^6 \sqrt{f(\rho_{*})}}\sqrt{\frac{L}{\Delta \rho_{*}}} P_s^2 \left(1+\frac{1}{3 f_{NL}} \right) \nonumber
\end{equation}
which clearly has dependence on $\Delta \rho_{*}$ in the denominator. Now for UV inflation we have $\rho_{*} >> \rho_0$ 
and therefore $f \sim 1$ to leading order. For a single warped throat we will find $\rho_{*} \le L$ since we assume that the gluing
region is located at $\rho \sim L$. This suggests that the maximal value for $\rho$ will be $L$. This also implies the constraint that
$L>> \Delta \rho_{*} $.
Now using (\ref{eq:identity}) we can solve for the brane tension in the expression above, and we find the solution
\begin{equation}
r < \frac{\kappa_{10}^2 R T_{4+2k}^2 v_{2k}^2 \pi^3 L^{4k+1}}{64 \Sigma_4 (\Delta N_{*})^6} \sqrt{\frac{L}{\Delta \rho_{*}}}P_s^2 \left(1+\frac{1}{3 f_{NL}} \right) 
\end{equation}
The KK-radius is constrained by the relation \cite{Davis:2010it}
\begin{equation}
R^2 = \frac{4 L^3}{9 \rho_0}
\end{equation}
where $L$ is the flux parameter, dependent on the string coupling and the flux integer. Solving for the various parameters in the above expression, and 
assuming that $\Delta N_{*} \sim 1$, we obtain the leading order bound
\begin{equation}
r < \frac{v_{2k}^2}{3 \Sigma_4} \frac{\pi^{3-8k/3}}{2^{7+4k}}\sqrt{\frac{l_s^2}{\rho_0 \Delta \rho_{*}}} (g_s M)^{(1+4k/3)} P_s^2
\end{equation}
after dropping $f_{NL}$. Let us assume that the four-cycle volume is given by $8\pi^2/3$, and the two-cycle volume by $4 \pi$. Using the WMAP 7 year
data \cite{Larson:2010gs, Komatsu:2010fb}, 
the normalisation for the scalar perturbation is $P_s^2 \sim 2.43 \pm 0.11 \times 10^{-9}$, and the bound on $r$ is $r < 0.24$. This allows
us to estimate the magnitude of the parameters in the bound for the three different cases of interest. The results are;
\begin{eqnarray}
(g_s M) \left(\frac{l_s}{\sqrt{\rho_0 \Delta \rho_{*}}}\right) &>& 6 \times 10^{10} \hspace{1cm} k=0 \nonumber \\
(g_s M) \left(\frac{l_s}{\sqrt{\rho_0\Delta \rho_{*}}}\right)^{3/7} & > &4 \times 10^4 \hspace{1.15cm} k=1 \\
(g_s M) \left(\frac{l_s}{\sqrt{\rho_0 \Delta \rho_{*}}}\right)^{3/11} & > &3 \times 10^3 \hspace{1.15cm} k=2 \nonumber
\end{eqnarray}
the DBI action is only valid for scales larger than the string scale, therefore we must have $\rho_0, \Delta \rho_{*} \ge l_s $ which acts to
suppress the flux term even for $\rho_0 \sim \mathcal{O}(1) l_s$. This multiplier term clearly has its maximum (of unity) when $\rho_0 = \Delta \rho_{*}=l_s$,
which means that the above constraints are then to be applied only to the product $g_s M$.
In general we see that as $ \rho_{0} \Delta \rho_{*}$ increases, the background flux must increase accordingly putting the 
theory under severe pressure for $k=0$. The $k=1, 2$ cases, corresponding to a wrapped $D6$ and $D8$-branes, could satisfy the bounds with sufficient fine tuning. 
The cosmology of such a configuration is left for future study.

One can ask whether there is also a lower bound on $r$ \cite{Kobayashi:2007hm, Lidsey:2007gq}. To see that this is also the case, we use the relation
\begin{equation}
1-n_s =  4\epsilon + \frac{2s}{1-\gamma^2}-\frac{2 M_p^2}{\gamma}\frac{T' H''}{TH}
\end{equation}
where primes denote derivatives with respect to $\phi$. Using the definition of the $f_{NL}$ parameter to eliminate the $\gamma^2$ terms, and using 
the continuity equation $\dot{\rho} = -3 H(P + \rho)$ in the final term above, we see that this expression becomes
\begin{equation}
1-n_s = \frac{r\sqrt{1+3 f_{NL}}}{4} - \frac{2s}{3 f_{NL}} + \frac{\dot{T}}{HT}.
\end{equation}
The expression above can be minimised provided that $\dot{T} \le 0$, which implies that
\begin{equation}
0 \ge \dot{\rho}\left(\frac{\rho^3(6+k)-\rho_0^3(3+k)}{2 \rho^3 \sqrt{f}} \right)
\end{equation}
however the term in parenthesis is positive definite $\forall k, \rho \ge \rho_0$ and therefore we find $\dot{\rho} \le 0$ which implies
that $\rho$ is rolling towards the tip. Under the assumption of large $f_{NL}$ and neglecting the slow roll parameters, this implies there
is a bound on $r$ such that
\begin{equation}
r \ge \frac{4 (1-n_s)}{\sqrt{1+3 f_{NL}}}
\end{equation}
which is the same as the bound on the (standard) type IIB theory. Using the (weak) bound that $|f_{NL}| \le 300$ and the WMAP result
for $n_s = 0.963$ \cite{Larson:2010gs, Komatsu:2010fb} we see that for the bounds on $r$ to be consistent we require
\begin{equation}
\frac{v_{2k}^2}{\Sigma_4} \frac{(g_s M)^{(5+8k)/6}}{2^k \pi^{8k/3}}\sqrt{\frac{l_s}{\rho_0}} >> 3.5 \times 10^7
\end{equation}
which we can translate into bounds on the flux term in the usual manner
\begin{eqnarray}\label{eq:minimal_bound}
(g_s M)\left( \frac{l_s}{\rho_0}\right)^{3/5} &>& 2 \times 10^{7} \hspace{1cm} k=0 \nonumber \\
(g_s M) \left(\frac{l_s}{\rho_0}\right)^{3/13} &> &7 \times 10^3 \hspace{1cm} k=1 \\
(g_s M) \left(\frac{l_s}{\rho_0}\right)^{1/7} &> &2 \times 10^2 \hspace{1cm} k=2 \nonumber
\end{eqnarray}
recall that this is the \emph{minimum} value necessary to be consistent with the lower bound. The bound for the $D8$-brane is 
by far the easiest to satisfy, indicating that these bounds are significantly relaxed for higher dimensional (wrapped) brane models. 

Let us now cross-check with the strong coupling bound, which has $\lambda << M^{3/2}$. Upon substitution of the string theory
parameters into the definition of the coupling, one can write the strong coupling constraint condition as a constraint on $g_s M$;
\begin{equation}
\frac{9 \pi}{4}\frac{\rho_0}{l_s} \frac{1}{M^{1/3}} << (g_s M) << \frac{9 \pi}{4} \frac{\rho_0}{l_s} M^2
\end{equation}
such a condition can be satisfied provided that the hierarchy between $\rho_0$ and $l_s$ is not too large (for fixed $M$). As one
increases the flux, the bound on the hierarchy is substantially weakened allowing for larger values of $\rho_0$ ensuring the validity
of the effective action.
\section{Discussion}
We have considered a simple extension of the recently proposed twisted inflation where we include higher dimensional branes wrapping compact cycles in the space $S^1 \times \mathcal{M}_4$.
In the relativistic limit, and for UV inflation, we have argued that the scalar-tensor ratio is bounded from above (like the models in type IIB). The lower
bound in our model is the same as that in DBI-inflation, suggesting that it is only dependent on the algebraic structure of the action. The upper
bound on $r$ in our model is, however, different from that in type IIB.
The main difference between these two theories is that in type IIA, our brane always wraps a compact $S^1$ associated with the KK-direction
in the field theory. Our results impose stringent bounds on the allowed fluxes of each solution, in order to satisfy the WMAP constraints. Unlike the type IIB
case, we cannot relate these fluxes to the Euler number of the Calabi-Yau space. Our results suggest that a modified version of $D8$-brane inflation
could lead to interesting cosmology in such a framework. It would be informative to consider the wrapped brane scenario in the slow roll regime, since 
this is likely to have an exact dual interpretation from the field theory perspective. Moreover, we could consider multiple KK-directions in the field theory
which correspond to torii in the string theory picture.
\begin{center}
\bf{Acknowledgments}
\end{center}
The work of JW is supported in part by NSERC of Canada.

\end{document}